\documentclass[nodate,showpacs,preprintnumbers,amsmath,amssymb,aps,prl]{revtex4-1}

\usepackage{color}

\usepackage{graphics} 
\newcommand*{\eps}{{\rlap{\lower2ex\hbox{$\,\,\tilde{}$}}{\epsilon_{ijk}}}}
\newcommand*{\EPS}{{\rlap{\lower2ex\hbox{$\,\,\tilde{}$}}{\epsilon_{i'j'k'}}}}
\newcommand*{\lmq}{{\rlap{\lower2ex\hbox{$\,\,\tilde{}$}}{\epsilon_{lmq}}}}
\newcommand*{\jmq}{{\rlap{\lower2ex\hbox{$\,\,\tilde{}$}}{\epsilon_{jmq}}}}
\newcommand*{\jql}{{\rlap{\lower2ex\hbox{$\,\,\tilde{}$}}{\epsilon_{jql}}}}
\newcommand*{\jlm}{{\rlap{\lower2ex\hbox{$\,\,\tilde{}$}}{\epsilon_{jlm}}}}
\newcommand*{\imq}{{\rlap{\lower2ex\hbox{$\,\,\tilde{}$}}{\epsilon_{imq}}}}
\newcommand*{\iql}{{\rlap{\lower2ex\hbox{$\,\,\tilde{}$}}{\epsilon_{iql}}}}
\newcommand*{\ilm}{{\rlap{\lower2ex\hbox{$\,\,\tilde{}$}}{\epsilon_{ilm}}}}
\newcommand*{\lmn}{{\rlap{\lower2ex\hbox{$\,\,\tilde{}$}}{\epsilon_{lmn}}}}
\newcommand*{\abc}{{\rlap{\lower2ex\hbox{$\,\,\tilde{}$}}{\epsilon_{abc}}}}
\newcommand*{\N}{{\rlap{\lower2ex\hbox{$\,\,\tilde{}$}}{N}}}
\newcommand{\tN}{{\rlap{\lower2ex\hbox{$\,\,\tilde{}$}}{N}}}
\newcommand*{\tM}{{\rlap{\lower2ex\hbox{$\,\,\tilde{}$}}{M}}}
\newcommand*{\imn}{{\rlap{\lower2ex\hbox{$\,\,\tilde{}$}}{\epsilon_{imn}}}}
\newcommand*{\qt}{\ln q^{\frac{1}{3}}}

\begin{document}

\title{Intrinsic Time Quantum Gravity}
\author{Hoi-Lai Yu}\email{hlyu@phys.sinica.edu.tw}
\affiliation{Institute of Physics, Academia Sinica, Taiwan}

\begin{abstract}
Correct identification of the true gauge symmetry of General Relativity being 3d spatial diffeomorphism invariant(3dDI) (not the conventional infinite tensor product group with principle fibre bundle structure), together with intrinsic time extracted from clean decomposition of the canonical structure yields a self-consistent theory of quantum gravity.  A new set of fundamental commutation relations is also presented. The basic variables are the 8 components of the unimodular part of the spatial dreibein and 8 $SU(3)$ generators which correspond to Klauder's momentric variables that characterize a free theory of quantum gravity.
The commutation relations are not canonical, but have well defined group theoretical meanings. All fundamental entities are dimensionless; and the quantum wave functionals are preferentially in the dreibein representation. The successful quantum theory of gravity involves only broad spectrum of knowledge and deep insights but no exotic idea.
\\
\\
Invited talk at $2^{nd}$ LeCosPA Symposium, {\it Everything about gravity}\\
 Taipei, $16^{th}$ Dec. 2015, contribution to the Proceedings
\end{abstract}


\maketitle
\section{Intrinsic Time and physics of the Hamiltonian constraint}

We had shown in\cite{SOOYU, SOOYU1} that the symplectic potential 
\begin{equation}\int\tilde{\pi}^{ij}\delta
q_{ij}=\int\bar{\pi}^{ij}\delta\bar{q}_{ij}+{\tilde\pi}\delta\ln
q^{\frac{1}{3}},\nonumber
\end{equation} can be cleanly separated into the conjugate pair, $(\ln q^{\frac{1}{3}}, {\tilde\pi})$, consisting of (one-third of) the logarithm of the determinant of the spatial metric and the trace of the momentum, from $({\bar q}_{ij}, {\bar \pi}^{ij})$, the unimodular part of the spatial metric with traceless conjugate momentum that allows a deparametrization of the theory wherein $\ln q^{\frac{1}{3}}$  plays the role of the intrinsic time variable for $\beta^2 = l-\frac{1}{3}>0$ with $l$ being the deformation parameter in the  DeWitt supermetric; $G_{ijkl}=\frac{1}{2}\left(q_{ik}q_{jl}+q_{il}q_{jk}\right)-l
q_{ij}q_{kl}$. This decomposition and identification of the intrinsic time variable point to a paradigm shift in the symmetries of Gravitation/space-time. 

The fundamental symmetries of Einstein's General Relativity(GR) can be revealed by carrying out its full canonical analysis.
Since space-time is not inert in Gravitation, the canonical analysis which has to be done without a fixed background had to wait until the work of Dirac\cite{Dirac2} and Arnowitt-Deser-Misner. The 4-d metric and the action can be expressed, essentially without loss of generality as,
\begin{equation}
ds^{2}= g_{\mu\nu}dx^{\mu}dx^\nu=
-N^2dt^2+q_{ij}(dx^i+N^idt)(dx^j+N^jdt),\nonumber
\end{equation} 
\begin{equation} 
S = \int dtd^3x\,({\tilde \pi}^{ij}\frac{{dq}_{ij}}{dt}
-{N^i}{H_i}-{N}{H}) + \rm{boundary \,\,
term};\nonumber
\end{equation} 
wherein several important features are revealed: only the spatial or 3-metric $q_{ij}$ is dynamical with conjugate momentum ${\tilde \pi}^{ij}$, and  the fields ${N^i}$ and ${N}$ (which are respectively called the shift and lapse functions, and they parametrize how the same spatial point is deformed from one hypersurface to the next) play the role of
Lagrange multipliers associated, respectively, with the super-momentum ${H_i}$ and super-Hamiltonian ${H}$ constraints,
\begin{equation} 
{H_i} =
-2q_{ik}\nabla_j{{\tilde \pi}}^{jk} \,{( =0 )}, \nonumber
\end{equation} 
\begin{equation} 
{H} = \frac{2\kappa}{\sqrt{q}}[{G_{ijkl}}{\tilde \pi}^{ij}{\tilde \pi}^{kl}+V(q_{ij}) ] \,( =0 ), \nonumber
\end{equation} 
and for GR($l = \frac{1}{2}$ and $V({q}_{ij}) =- \frac{q}{(2\kappa)^2}[R - 2\Lambda_{\it{eff}} ]$).
By the usual method of counting in canonical analysis, there are thus 6 degrees of freedom (d.o.f.) associated with the canonical pairs $(q_{ij},{\tilde \pi}^{ij}$) for GR in four-dimensions, subject to 4 constraints, resulting in 2 net field d.o.f. in the theory.  The constraints and the algebra, or commutation relations they obey  (which for GR is also called the Dirac algebra)  also shed light on the fundamental symmetries of the theory. Under the action of the super-momentum constraint (also called diffeomorphism constraint for reasons that will soon become clear), the fundamental variables changes by
\begin{equation}
 \{f(q_{ij}, {\tilde\pi}^{ij}), H_k[N^k]\} _{\rm P.B.}={\cal L}_{\vec N}f , \qquad H_i[N^i] \equiv \int
N^iH_i d^3x;\nonumber
\end{equation}
 i.e. by a diffeomorphism.

So although it is possible to interpret the constraint $H_i$ causes the same effect on the fundamental variables as the change induced by an infinitesimal general coordinate transformation, it should be emphasized that, 
1)the true symmetry of the theory is dictated by the form of the constraints and the precise transformations they generate, and not by changes in the integration dummy spatial coordinate variable which is integrated over in both the action and in the Hamiltonian;
2)$H_i$ generates diffeomorphisms which are changes in the dynamical fields evaluated at the {\it same} coordinate point, for instance, $\delta_{\vec N}q_{ij}(x) = \{q_{ij}(x), H_k[N^k]\} _{P.B.}= {\cal L}_{\vec N}q_{ij}(x)$, and in this aspect there is no distinction between usual Yang-Mills gauge transformations which are deemed `internal' while GR is often naively associated with `external' or space-time {\it coordinate} transformations. 
What is important is that the symmetries are not realized through   `transformation of coordinates' but through the transformation of fields in which the coordinate labels are inert and play no active role. Thus as far as spatial diffeomorphisms are concerned, they are genuine {\it gauge symmetries} of Einstein's theory but with spatial metric being the basic variable and are on equal footing with usual Yang-Mills `internal' gauge symmetries which describes symmetries under changes at the same coordinate point with connection variables.

For GR, the interpretation of the changes generated by the super-Hamiltonian or Wheeler-DeWitt constraint, $H$, is not so straightforward. It can be shown that
\begin{eqnarray} 
\delta_{N}q_{ij} &=& \{q_{ij}, H[N]\} =  \frac{4N\kappa}{\sqrt q}G_{ijkl}{\tilde \pi}^{kl}={\cal L}_{Nn^\mu}q_{ij} \,[{\rm modulo.\, EOM}], \nonumber \\
\delta_{N}{\tilde \pi}^{ij} &=& \{{\tilde \pi}^{ij}, H[N]\}=\frac{N}{2} q^{ij} H -N{\sqrt q}(q^{ki}q^{lj} -q^{ij}q^{kl})R_{kl} + {\cal L}_{Nn^\mu}{\tilde \pi}^{ij}.\nonumber
\end{eqnarray}
Thus the constraint generates diffeomorphism or Lie derivative, ${\cal L}_{Nn^\mu}$, in the direction normal to the Cauchy surface {\it only  on-shell} on the constraint surface (i.e. when the constraints hold) and also when the equations of motion (EOM) are imposed (which for Einstein's theory in vacuum is the vanishing of the spacetime Ricci curvature tensor $R_{\mu\nu}$). This implies that in the quantum context, GR does not possess the full gauge symmetry of
four-dimensional diffeomorphisms, but spatial diffeomorphisms will remain intact. This is corroborated by the fact that the Dirac algebra is not the algebra of 4-dimensional diffeomorphisms. The call to abandon 4-covariance is not new. In the simplification of the Hamiltonian analysis of GR, the fact that only the spatial metric is dynamical lead Dirac to conclude that `four-dimensional symmetry is not a fundamental property of the physical world'\cite{Dirac2}. In his seminal article, Wheeler had emphasized that space-time is a concept of `limited applicability', and it is 3-geometry, not 4-geometry, which is fundamental in Quantum Geometrodynamics\cite{Wheeler}.

In quantum field theories (QFT), true gauge symmetries have clear interpretation in the canonical context in Dirac quantization. Physical quantum states are  annihilated by the constraints, and this have the direct interpretation that the state is invariant under symmetry transformations of the fundamental configuration variable. For instance in Yang-Mills gauge theories, the Gauss Law constraint is
$G^b({\bf x})\Psi[A_{ia}] =\nabla_j{\hat \pi}^{jb}({\bf x})\Psi[A_{ia}]  =0$.  Realizing the conjugate momentum of $A_{ia}$ realized in the connection representation by operator, $ {\hat\pi}^{ia} = \frac{\hbar}{i}{{\delta }{\delta A_{ia}}}$ and Taylor expansion leads to
\begin{eqnarray}
\Psi[A_{ia}+ \delta_{gauge}A_{ia}] &=& \Psi[A_{ia} ] + \int ({\delta _{\rm gauge}A_{jb}}({\bf x})){\frac{\delta \Psi }{\delta A_{jb}}(x)}d^3{\bf x}  \nonumber \\
&= &\Psi[A_{ia}]+ \frac{i}{\hbar}(\int \eta_bG^b d^3{\bf x})\Psi [A_{ai}]=\Psi[A_{ia}],\nonumber
\end{eqnarray}
wherein ${\delta _{\rm gauge}A_{jb}}= -\nabla_j\eta_b$, together with integration by parts over compact Cauchy surface without boundary, or with assumed vanishing of gauge parameter $\eta_b$ on the boundary. This {\it linear dependents in canonical momentum}, manifests the {\it invariance under local gauge transformations} of all physical states in QFT. In precisely the same manner in GR, with regard to spatial diffeomorphisms\cite{YM},
\begin{equation}
\Psi[q_{ij}+ \delta_{\rm Diffeo.}q_{ij}] = \Psi[q_{ij} ] + \int ({\delta _{\rm Diffeo.}q_{ij}}){\frac{\delta \Psi }{\delta q_{ij}}}d^3{\bf x} =  \Psi[q_{ij} ] + \frac{i}{\hbar}(\int N^iH_i d^3{\bf x})\Psi[q_{ij}],\nonumber 
\end{equation}
since $\delta_{\rm Diffeo.}q_{ij} = {\cal L}_{\vec N}q_{ij} = \nabla_i N_j + \nabla_j N_i$, the momentum operator $ {\hat {\tilde \pi}}^{ij} = \frac{\hbar}{i}{\frac{\delta }{\delta q_{ij}}}$,  and the diffeomorphism constraint $H_i \Psi[q_{ij}]=0$.
In Table 1, we enlist the differences in gauge structures between 3dDI and conventional Yang-Mills gauge theories.

Now, one can easily see that the{ \it quadratic canonical momentum dependence} of the Hamiltonian constraint will deprive its role of generating temporal gauge transformations; but only constraints
\begin{equation}\label{Ham}
0\cong \frac{\sqrt{q}}{2\kappa} H = {\bar G}_{ijkl}{\bar \pi}^{ij}{\bar \pi}^{kl} -\beta^2{\tilde\pi}^2 + V(q_{ij})
=({\bar H}-\beta{\tilde\pi})({\bar H}+\beta{\tilde\pi})\Rightarrow \bar H=\pm\beta{\tilde\pi}; 
\end{equation}
wherein
$
\bar {H}(\bar{\pi}^{ij}, \bar {q}_{ij}, q)  := \sqrt{ \bar{G}_{ijkl}\bar{\pi}^{ij}\bar{\pi}^{kl} +  V(\bar {q}_{ij}, q) }$
plays the role of local Hamiltonian density that generates real intrinsic time evolution.  Although the constraints form a first class algebra, and the lapse function $N$ is {\it a priori} arbitrary; classically, the physical meaning of $N$ is fixed {\it a posteriori} by the EOM and constraints\cite{SOOYU}, 
\begin{equation}
N =\frac{\sqrt{q}(\partial_t \ln q^{1/3} -\frac{2}{3}\nabla_iN^i)}{4\beta\kappa {\bar H}}.\nonumber  
\end{equation}
Paradigm shift in the role of the Hamiltonian constraint and the identification of intrinsic time variable from the spatial metric collude to yield a theory of quantum gravity with underlying 3dDI gauge symmetry dictated by the spatial metric. 
\begin{table}
 \caption{Comparison of the two different gauge structures}
{\begin{tabular}{@{}cccc@{}}
\toprule
& Diffeomorphism Gauge Structures & Yang-Mills Gauge Structures 
\\\colrule
Basic Variables\hphantom{00} & \hphantom{0} Spatial metric tensor $q_{ij}$ & \hphantom{0} Gauge connection $A_{ia}$  \\
Symmetry Generators\hphantom{00} & \hphantom{0}$H_{i}({\bf x})=-2q_{ik}\nabla_{j}\pi^{jk}({\bf x})(=0)$ & \hphantom{0}$G^{a}({\bf x})=\nabla_{i}\pi^{ia}({\bf x})(=0)$ \\
Gauge transformation\hphantom{00} & \hphantom{0}$[q_{ij}({\bf x}), H_{k}[N^{k}]]={\cal L}_{ \vec{N}}q_{ij}({\bf x})$; & \hphantom{0}$[A_{ia}({\bf x}),G^{b}[\eta_{b}]]= -\nabla_{i}\eta_{a}({\bf x})G^{b}[\eta_{b}]$; \\
\hphantom{00} & \hphantom{0} $H_{i}[N^{i}]=\int N^{i}H_{i} d^3{\bf x}$& \hphantom{0}$ G^{b}[\eta_{b}]=\int \eta_{b}G^{b} d^3{\bf x}$ \\
Commutation Relations \hphantom{00} & \hphantom{0} $[H_i ({\bf x}),H_j ({\bf y})]$& \hphantom{0} $[G^{a}({\bf x}), G^{b}({\bf y})] $ \\
\hphantom{00} &  \hphantom{0} $= H_j ({\bf x})\partial_i \delta({\bf x}-{\bf y})+ H_i ({\bf y})\partial_j \delta ({\bf x}-{\bf y})$ & \hphantom{0} $=i f^{ab}_{\,\,\,\,\,c} G^{c}({\bf x})\delta({\bf x}-{\bf y})$\\
Potentials \hphantom{00} & \hphantom{0} $V\sim[\frac{\delta exp(CS)}{\delta q_{ij}}]^{2}$& \hphantom{0} $V\sim[\frac{\delta exp(CS)}{\delta A_{ia}}]^{2}$ \\
\hphantom{0} & Not product of identical group(i.e. $SL(3R)$)  & Infinite tensor product group \hphantom{0} $\prod_{x}G$; &  \\ Locality \& Dimension \hphantom{0} & \hphantom{0} at each spatial point of base manifold &$G$=finite dimensional Lie group \\
\hphantom{0} & i.e. not of principle fibre structure \hphantom{0} & =usually referred to as the `gauge group'  \\
 \\\botrule
\end{tabular}}
\end{table}
The physical contents of this new theory are equivalent to GR\cite{SOO}, however, with the great advantages of being capable of modifying the potential, i.e. to include the conformal structures through the Cotton-York tensor to achieve renormalizability of the theory  without encountering any inconsistency in the constraint algebra.

Before passing, note that while $q$ is a tensor density, the multi-fingered intrinsic time interval,  $\delta\qt = \frac{q^{ij}}{3}\delta q_{ij}$, is a scalar entity although being non-integrable. Hodge decomposition for any compact Riemannian manifold without boundary yields,
$\delta \ln q^{\frac{1}{3}} =\delta{T}+\nabla_i\delta{Y}^i$, wherein the integrable gauge-invariant part of $\delta\qt$ is $\delta T =\frac{2}{3}\delta \ln V_{\rm spatial}$ proportional to the 3dDI logarithmic change in the spatial volume\cite{SOO}. Upon quantization in \eqref{Ham}, the Schrodinger equation can be derived, 
 \begin{equation}
 i\hbar\frac{\delta \Psi}{\delta T} = \int i\hbar\frac{\delta \Psi}{\delta\qt(x) }\frac{\delta \qt(x)}{\delta T}d^3x=\left[\int\frac{{\bar H}(x)}{\beta}d^3x\right]\Psi ={H}_{\rm Phys}\Psi,\nonumber
 \end{equation}
$H_{\rm Phys}$ is the physical Hamiltonian generating evolution in global intrinsic time $T$.

 \section{Momentric and Commutation relations of geometrodynamics}
 
In usual quantum theories, the fundamental {\it canonical} CR, $[Q({\bf x}), P({\bf y})] = i\hbar\delta({\bf x}-{\bf y})$, implies $\frac{P}{\hbar}$ is the generator of translations of $Q$ which are also symmetries of the `free theories' in the limit of vanishing interaction potentials.
For geometrodynamics, the corresponding canonical CR is $[q_{ij}({\bf x}), {\tilde\pi}^{kl}({\bf y})] = i\hbar\frac{1}{2}(\delta^k_i\delta^l_j + \delta^l_i\delta^k_j)\delta({\bf x}-{\bf y})$. However, neither positivity of the spatial metric is preserved under arbitrary translations in superspace generated by the conjugate momentum; nor is the `free theory' invariant under  translations, because the kinetic part, $G_{klmn}{\tilde\pi}^{kl}{\tilde\pi}^{mn}$, of the Hamiltonian dependents on $q_{ij}$. In quantum gravity, states which are infinitely peaked at the flat metric, or for that matter any particular metric with its corresponding isometries, cannot be postulated ad hoc; consequently, the underlying symmetry of even the `free theory' is obscure.
The Poisson brackets for the barred variables are,
\begin{eqnarray}
\label{qT}
\{ \bar{q}_{ij}({\bf x}),\bar{q}_{kl}({\bf y})\} &=&0,\,\,\,\,\,\, \{\bar{q}_{kl}({\bf x}),{\bar{\pi}}^{ij}({\bf y})\}= P^{ij}_{kl}\,\delta({\bf x}-{\bf y}),\nonumber \\
\{ \bar{\pi}^{ij}({\bf x}),\bar{\pi}^{kl}({\bf y})\} &=& \frac{1}{3}({\bar q}^{kl}{\bar{\pi}}^{ij} -{\bar q}^{ij}{\bar{\pi}}^{kl})\delta({\bf x}-{\bf y});
\end{eqnarray}
with $P^{ij}_{kl} :=  \frac{1}{2}(\delta^i_k\delta^j_l + \delta^i_l\delta^j_k) - \frac{1}{3}\bar{q}^{ij}\bar{q}_{kl}$ denoting the traceless projection operator. This set is not strictly canonical.
In the metric representation, the implementation of ${\bar{\pi}}^{kl}$ as traceless, symmetric, and self-adjoint operators is problematic. Remarkably, these difficulties can be cured by passing to the `momentric variable' (first introduced by Klauder\cite{Klauder}) which is classically $\bar \pi^{i}_{j} = \bar q_{jm}\bar \pi^{im}$.
In terms of spatial metric and momentric variables, the fundamental CR postulated (from which the classical Poisson brackets corresponding to \eqref{qT} can be recovered)  are then\cite{ITQG}
\begin{eqnarray}
\label{REL}
[ \bar q_{ij}({\bf x}), \bar q_{kl}({\bf y})]&=&0,\,\,\,\,\,\,\,\,\,\,
[\bar q_{ij}({\bf x}), {\bar{\pi}}^{k}_{l}({\bf y})]= i\hbar\bar{E}^k_{l(ij)}\delta({\bf x}-{\bf y}),\nonumber \\
\bigl[ {\bar \pi}^{i}_{j}({\bf x}), {\bar \pi}^{k}_{l}({\bf y})\bigr] &=& \frac{i\hbar}{2}(\delta^k_j{\bar{\pi}}^i_l-\delta^i_l{\bar{\pi}}^k_j)\delta({\bf x}-{\bf y});
\end{eqnarray}\noindent
wherein $\bar{E}^i_{j(mn)}:=\frac{1}{2}(\delta^i_m\overline{q}_{jn}+\delta^i_n\overline{q}_{jm})-\frac{1}{3}\delta^i_j\overline{q}_{mn}$ (with
$\delta^{j}_{i}\bar{E}^i_{j(mn)} =$$ \bar{E}^i_{j(mn)} \bar q^{mn}=0$; $\bar{E}^i_{jil}=\bar{E}^i_{jli}=\frac{5}{3}\overline{q}_{jl}$) is the vielbein for the
supermetric ${\bar G}_{ijkl} =  \bar{E}^m_{n(ij)}\bar{E}^n_{m(kl)}$.
Quantum mechanically, the momentric operators and CR {\it can be explicitly realized  in the metric representation} by ${\hat {\bar{ \pi}}}^{i}_{j}({\bf x}):=\frac{\hbar}{i}\bar{E}^i_{j(mn)}({\bf x})\frac{\delta}{\delta \bar q_{mn}({\bf x})} =\frac{\hbar}{i}\frac{\delta}{\delta \bar q_{mn}({\bf x})}\bar{E}^i_{j(mn)}({\bf x})=\hat{\bar{ \pi}}^{\dagger i}_{j}({\bf x})$ which are self-adjoint on account of $[\frac{\delta}{\delta\bar{q}_{mn}({\bf x})},\bar{E}^i_{j(mn)}({\bf x})]=0$. 

\section{New commutation relations for quantum gravity}

Momentric variables, ${\hat {\bar{ \pi}}}^{i}_{j}$, generate $SL(3,R)$ transformations  of $\bar{q}_{ij} =\delta^{ab}{\bar e}_{ai}{\bar e}_{bj} $ which preserve its positivity and unimodularity. Moreover, they generate at each spatial point, an $SU(3)$ algebra. In fact, with $3\times 3$ Gell-Mann matrices $\lambda^{A=1,...,8}$,  it can be checked that $T^{A}({\bf x}):= \frac{1}{\hbar\delta({\bf 0})}(\lambda^{A})^{j}_{i}\hat{\bar \pi}^{i}_{j}({\bf x})$  generates the $SU(3)$ algebra with structure constants $f^{AB}\,_C$.
The 5 verse 8 asymmetry in the independent components between  ${\bar q}_{ij}$ and the symmetric traceless, $\tilde \pi^{ij}$(also $\bar \pi^{i}_{j}$) in \eqref{qT} and \eqref{REL}  is rectified by the unimodular dreibein-traceless momentric variables, $({\bar e}_{ai} := e^{-\frac{1}{3}}e_{ai}, T^A)$; each having 8 independent components and obey the {\it  advocated new  fundamental CR  }\cite{NCR},
\begin{eqnarray}\label{fundamental}
\bigl[{\bar e}_{ai}({\bf x}), {\bar e}_{bj}({\bf y})\bigr] &=&0,\,\,\,\, \bigl [{\bar e}_{ai}({\bf x}),T^A({\bf y})\bigr] =i(\frac{\lambda^A}{2})^k_i{\bar e}_{ak}\frac{\delta({\bf x}-{\bf y})}{\delta({\bf 0})},\nonumber \\  
\bigl[T^{A}({\bf x}),T^{B}({\bf y})\bigr]&=& i{f}^{AB}_CT^{C}\frac{\delta({\bf x}-{\bf y})}{\delta({\bf 0})}.
\end{eqnarray}
A  number of intriguing features are encoded in this set of CR.
Because of \eqref{fundamental}, the quantum wave functionals are preferentially selected to be in the dreibein representation. It is noteworthy that  all entities in \eqref{fundamental}, including $({\bar e}_{ai}, T^A)$ and $\frac{\delta({\bf x}-{\bf y})}{\delta({\bf 0})}$,  are dimensionless; and {\it neither the gravitational coupling constant nor Planck's constant make their appearance}. ${\delta({\bf 0})}:=\lim_{x\rightarrow y} \delta({\bf x}-{\bf y})$ denotes the coincident limit; so there are no divergences in $\frac{\delta({\bf x}-{\bf y})}{\delta({\bf 0})}$ which is unity in the coincident limit and vanishing otherwise.
The second CR in \eqref{fundamental} implies
$
\exp(i\int\alpha_BT^B\delta({\bf 0})d^3{\bf y}'){\bar e}_{ai}({\bf x})\exp(-i\int\alpha_AT^A\delta({\bf 0})d^3{\bf y}) = (\exp({\frac{\alpha_A({\bf x})\lambda^A}{2}}))^j_i{\bar e}_{ja}({\bf x}),
$
with $\exp({\frac{\alpha_A({\bf x})\lambda^A}{2}})$ being a local $SL(3,R)$ transformation(not gauged) which will break down to the 3dDI(gauged) subgroup when interactions are introduced. $SL(3,R)/3dDI$ generates all 3dDI inequivalent quantum states from any 3dDI initial state.
While the  CR in \eqref{fundamental} states that $T^A({\bf x})$  generates, at each spatial point, a separate $SU(3)$ algebra that characterize a free quantum theory.  Further details on intrinsic time quantum geometrodynamics and discussions on the causality and time-ordering can be found in  
Chopin Soo's plenary talk in this meeting. 
\section*{Acknowledgments}
 I would like to thank Eyo Eyo Ita III
and Chopin Soo for beneficial discussions.

\bibliographystyle{ws-procs961x669}
\bibliography{ws-pro-sample}

\end{document}